\documentclass[conference]{IEEEtran}

\usepackage[pdftex]{graphicx}

\usepackage[hide]{chato-notes}

\usepackage{flushend}
\usepackage{booktabs}
\usepackage{url}
\usepackage{caption}
\usepackage{subcaption}
\usepackage{amssymb}
\usepackage{balance}
\usepackage{xcolor}
\usepackage{booktabs}
\usepackage{multirow}

\usepackage{optidef}
\usepackage{amsthm}
\usepackage{algpseudocode}
\usepackage{algorithm}
\usepackage{listings}
\usepackage[symbol]{footmisc}

\usepackage{breqn}

\DeclareMathOperator*{\argmin}{argmin}

\newcommand{\our}{\textsc{DecWVC}\xspace}
\newcommand{\ours}{\our}
\newcommand{\ad}{$\mathcal{D}$\xspace}

\newcommand{\problem}{\textsc{WMVC}\xspace}
\newcommand{\fastvc}{\textsc{FastWVC}\xspace}
\newcommand{\apw}{\textsc{APW}\xspace}
\newcommand{\timemetric}{\textsc{ExTime}\xspace}
\newcommand{\mpn}{\textsc{MPN}\xspace}
\newcommand{\numrealgraphs}{20\xspace}

\newcommand{\barabasi}{scale-free\xspace}
\newcommand{\randg}{random\xspace}
\newcommand{\nws}{small-world\xspace}

\newcommand{\selection}{\lstinline{Selection()}\xspace}
\newcommand{\optimize}{\lstinline{Optimize()}\xspace}

\newcommand{\cutoff}{100\xspace}

\newcommand\CONDITION[2]%
   {\begin{tabular}[t]{@{}l@{}}#1#2\end{tabular}}
\algdef{SE}[WHILE]{While}{EndWhile}[1]%
   {\algorithmicwhile\ \CONDITION{#1}{\ \algorithmicdo}}%
   {\algorithmicend\ \algorithmicwhile}
\algdef{SE}[FOR]{For}{EndFor}[1]%
   {\algorithmicfor\ \CONDITION{#1}{\ \algorithmicdo}}%
   {\algorithmicend\ \algorithmicfor}
\algdef{S}[FOR]{ForAll}[1]%
   {\algorithmicforall\ \CONDITION{#1}{\ \algorithmicdo}}
\algdef{SE}[REPEAT]{Repeat}{Until}%
   {\algorithmicrepeat}[1]%
   {\algorithmicuntil\ \CONDITION{#1}{}}
\algdef{SE}[IF]{If}{EndIf}[1]%
   {\algorithmicif\ \CONDITION{#1}{\ \algorithmicthen}}%
   {\algorithmicend\ \algorithmicif}
\algdef{C}[IF]{IF}{ElsIf}[1]%
   {\algorithmicelse\ \algorithmicif\ \CONDITION{#1}{\ \algorithmicthen}}
\makeatletter

\begin{document}

\title{A Fast-Converging Decentralized Approach to the Weighted Minimum Vertex Cover Problem}

\author{\IEEEauthorblockN{Matteo Mordacchini}
 \IEEEauthorblockA{National Research Council of Italy, Italy\\
 Email: matteo.mordacchini@iit.cnr.it}
 \and
 \IEEEauthorblockN{Emanuele Carlini}
 \IEEEauthorblockA{National Research Council of Italy, Italy\\
 Email: emanuele.carlini@isti.cnr.it}
 \and
 \IEEEauthorblockN{Patrizio Dazzi}
 \IEEEauthorblockA{University of Pisa, Italy\\
 Email: patrizio.dazzi@unipi.it}}

% \author{\IEEEauthorblockN{Michael Shell}
% \IEEEauthorblockA{School of Electrical and\\Computer Engineering\\
% Georgia Institute of Technology\\
% Atlanta, Georgia 30332--0250\\
% Email: http://www.michaelshell.org/contact.html}
% \and
% \IEEEauthorblockN{Homer Simpson}
% \IEEEauthorblockA{Twentieth Century Fox\\
% Email: homer@thesimpsons.com}
% \and
% \IEEEauthorblockN{James Kirk\\ and Montgomery Scott}
% \IEEEauthorblockA{Starfleet Academy\\
% San Francisco, California 96678--2391\\
% Telephone: (800) 555--1212\\

\maketitle

\begin{abstract}
%The weighted minimum vertex cover is a classical optimization problem where, given a graph, the objective is to find the smallest subset of vertices that cover all edges while minimizing the total weight. 
%
%This problem is NP-hard, and approximation methods are commonly employed for large graphs.
%
%This paper introduces \our, an approximate, decentralized, iterative, and message-passing algorithm to solve the weighted minimum vertex cover problem. 
%
%The algorithm relies solely on the local information of nodes (i.e., the 1-hop neighborhood), making it suitable for networks where entities need to autonomously determine their inclusion in the vertex cover, such as in multi-agent and peer-to-peer networks.
%
%The approach has been thoroughly evaluated with a heterogeneous mix of graphs, both real and synthetic. 
%
%The evaluation showed that the approach performs well in \barabasi networks, and due to its convergence speed and minimal computational footprint, it also performs quickly as an in-memory sequential solver.
%
%Moreover, the vertex cover solution is, on average, within 1\%, and in some cases, even surpasses that of a centralized state-of-the-art approach.

We address the problem of computing a Minimum Weighted Vertex Cover (MWVC) in a decentralized network. MWVC, a classical NP-hard problem, is foundational in applications such as network monitoring and resource placement. We propose a fully decentralized protocol where each node makes decisions using only local knowledge and communicates with its neighbors. The method is adaptive, communication-efficient, and avoids centralized coordination. We evaluate the protocol on real-world and synthetic graphs, comparing it to both centralized and decentralized baselines. Our results demonstrate competitive solution quality with reduced communication overhead, highlighting the feasibility of MWVC computation in decentralized environments.
\end{abstract}


\IEEEpeerreviewmaketitle

\section{Introduction}

%\todo{guardarsi il CONGEST e il LOCAL distribution models}

% \todo{we still don't have the results from the silly heuristics!!!}

In this work, we study the problem of computing a Minimum Weighted Vertex Cover (MWVC) in a fully decentralized setting. Each node knows only its weight and the identities of its neighbors. MWVC is a fundamental optimization problem~\cite{Dinur2005439} in network design, with applications ranging from resource placement to intrusion detection and distributed monitoring~\cite{10296816,makris2024pro, gottlieb2014efficient,makris2024optimizing}.
Centralized MWVC algorithms are often infeasible in large-scale or privacy-sensitive settings. In contrast, a decentralized approach enables nodes to operate independently with limited coordination~\cite{sac25}, making the system scalable and resilient to single points of failure. However, achieving good approximation quality in such settings remains challenging.
%, including in the weighted case, where each node has a different cost.
We propose a decentralised simple, local protocol based on iterative updates and local decision rules. Nodes exchange limited information with their neighbors, collaborating to incrementally construct the cover set.
%converge to a near-optimal solution. 
Our method is robust to changes in topology and node weight distributions, and its decentralised nature allows it to scale with the dimension of the network.  
%making it suitable for a wide range of practical scenarios.
We validate our approach using both synthetic and real-world graphs. The results show that our protocol achieves a high-quality vertex cover with low communication and computation overhead, 
%showing that \our solution is 
within 1\% in terms of the cover quality when compared to centralized state-of-the-art solutions.

In summary, our contribution can be summarized by these two important points: i) We propose a novel decentralized algorithm for the \problem computation, which relies only on local information and exhibits fast convergence and promising solution quality; ii) We present an extensive experimental evaluation, considering a heterogeneous set of networks, including synthetic sparse and dense, and real graphs.
% \begin{itemize}
%     \item We propose a novel decentralized algorithm for the \problem computation, which relies only on local information and exhibits fast convergence and promising solution quality;
%     \item We present an extensive experimental evaluation, considering a heterogeneous set of networks, including synthetic sparse and dense, and real graphs.\\
% \end{itemize}

The rest of this paper is structured as follows: Section~\ref{sec:related-work} summarizes the existing scientific literature of the \problem, focusing on distributed approaches.
In Section~\ref{sec:approach}, we present a formal definition of the WMVC problem and our proposed algorithm (\our).
%
%Section~\ref{sec:math} offers a comprehensive mathematical model of the devised solution. 
%
A comprehensive experimental evaluation that validates and compares \our approach is described in detail in Section~\ref{sec:exp}. 
Lastly, Section~\ref{sec:conclusion} summarizes the conclusions we have reached.

\section{Related Work}
\label{sec:related-work}

The weighted minimum vertex cover is a well-known NP-complete graph theory problem.
% and one of the 21 Karp's list of NP-complete problems. 
% %
% In its weighted version, the problem is to select a set of nodes that covers all edges and whose total weight is minimized. 
%
Due to its difficulty, solutions for the \problem are usually approximated for relatively large graphs. 
A categorization of the most recent approaches can be found in Chen et al. \cite{chen2024vertex}, while Taoka et al. provides \cite{taoka2012performance} an experimental comparison of different centralized approaches.
Popular centralized approaches solve the problem by building an initial set with nodes with a high probability of being in the optimal set and then extending that initial set to be a valid cover of the graph. This approach was first proposed by Cai et al. \cite{10.5555/2832249.2832353} for the unweighted minimum cover and then adapted to the weighted version in \cite{cai2018improving} and \fastvc \cite{CAI201964}. 
The FastWVC algorithm introduces two key strategies to enhance the efficiency of solving the MWVC problem on large-scale graphs. The first strategy aims to generate a high-quality initial vertex cover. This procedure selects vertices based on their weights and the number of uncovered edges connected to them, ensuring that the initial solution is effective and efficient. The second strategy involves a new exchange step for reconstructing the vertex cover. This step iteratively improves the solution by exchanging vertices in and out of the cover, optimizing the total weight while maintaining coverage.
%

% Li et al. \cite{li2016efficient} propose a similar solution that employs a dynamic weighting of the edges to drive the reconstruction of the vertex cover. 
% %
% Another notable approach from Langedal et al. \cite{langedal2022efficient} uses a Graph Neural Network to predict whether a node would be part of the initial set combined with a local search used to improve on the initial solution.

% %
% Other recent centralized approaches rely on evolutionary strategies. 
% Wang et al. \cite{wang2021fast} combine a master-apprentice evolutionary with a tabu search.
% Qiu et al. \cite{qiu2022population} integrate evolutionary strategy with game-theory to find a Nash equilibrium, ensuring no redundant nodes in the vertex cover.
% %
% Li et al. \cite{li2023evolutionary} propose a methodology based on an evolutionary algorithm to find an initial solution and then improve it using the snowdrift game, a well-known method borrowed from game theory.
% %

% Ghaffari et al. \cite{10.1145/3350755.3400260} propose an approach in which the vertexes are randomly distributed to a set of machines and perform parallel iterations on the subgraphs. After the iterations a number of vertexes are inserted in the cover. Then, communicate the results and repeat them for a number of steps. 
% %
% %
% Xu et al.~\cite{Xu} presented a new solver for the MWVC problem based on a novel reformulation to a series of SAT instances using a primal-dual approximation algorithm as a starting point.
%
%

Regarding distributed algorithms, several approaches have been proposed recently (see a summary in Table \ref{tab:sota}). Here, we focus on those approaches that consider the graph as the communication network, with each node acting as a separate agent whose only knowledge is its connections with neighboring nodes.
\begin{table}[t!]
    \footnotesize
    \centering
    \begin{tabular}{l|c|l}
        \toprule
         \textbf{Reference} &  \textbf{Year} & \textbf{Strategy} \\
         \midrule
         Grandoni et al. \cite{grandoni2008distributed} & 2008 & maximal matchings on micro-nodes\\
         \r{A}strand and Suomela \cite{10.1145/1810479.1810533} & 2010 & edge packing and graph colouring\\  
         Bar-Yehuda et al. \cite{bar2017distributed} & 2017 & local ratio\\ 
         Ben-basat et al. \cite{ben2018deterministic} & 2018 & local ratio\\
         Ghaffari et al. \cite{10.1145/3350755.3400260} & 2020 & parallel subgraphs\\
         Sun et al. \cite{sun2020distributed} & 2020 & game theory\\
%         Sun et al.\cite{sun2021better} & 2021 & game theory\\
         Faour et al. \cite{faour2021distributed} & 2021 & reduction to bipartite graph\\
         \bottomrule
    \end{tabular}
    \caption{Summary of selected recent distributed approaches for the weighted minimum vertex cover problem}
    \label{tab:sota}
\end{table}

An early approach from Grandoni et al. \cite{grandoni2008distributed} leverages a graph expansion technique to solve the \problem. In particular, it expands each node in a set of micro-nodes depending on the weight associated with the node. Then, a maximal matching in the auxiliary graph is computed, and the vertex cover is given by nodes for which all corresponding micro-nodes are matched. 
The solution from \r{A}strand and Suomela \cite{10.1145/1810479.1810533} augments the graph by extending the set of edges by building a maximal edge-packing and then uses graph coloring to identify nodes that belong to the cover.
Another solution that modifies the original graph has been recently proposed by Faour et al.\cite{faour2021distributed}. They leverage the properties of bipartite graphs to find the minimum cover set on such graphs. The approach for general graphs involves reducing the problem to bipartite graphs using a bipartite double cover technique and then translating the solution back to the original graph.

More recent solutions, notably Bar-Yehuda et al. \cite{bar2017distributed}, employed a \textit{local ratio} strategy to the distributed context. The local ratio repeatedly reduces weights from both endpoints of an edge without going below zero. The set of vertices with no remaining weight forms a (2)-approximation for \problem.
Ben-basat et al. \cite{ben2018deterministic} improve over the local ratio technique of \cite{bar2017distributed}, providing a generalization that reduces the number of rounds to reach convergence.
Other recent solutions explored game theoretical approaches for the vertex cover. 
In the paper from Sun et al. \cite{sun2020distributed}, each node in the network starts with a random action, either 0 (not in the cover) or 1 (in the cover). 
At each round, by communicating with its neighbors, each node changes the action to the one that maximizes the utility toward the vertex cover, also considering 
%past actions. In doing so, nodes also consider 
their regret value, which measures the potential improvement in utility if the node switches to its best response.
%
%Sun et al. \cite{sun2021better} improve on the former by differentiating between special and normal nodes. Special nodes have their weight greater than the sum of the weights of their neighbors that are not selected, and their response is perturbed to converge faster to a Nash equilibrium.

\section{Decentralized Weighted Minimum Vertex Cover}
\label{sec:approach}

% \subsection{Preliminaries}\label{sec:preliminaries}
% We consider an undirected graph \( G = (V, E) \), where \( V \) is the set of vertices and \( E \) is the set of edges connecting pairs of vertices in \( V \). Each vertex \( i \in V \) has an associated positive {\em weight} defined by the function \( w: V \to \mathbb{R}^+ \), where \( w(i) \) denotes the weight of vertex \( i \). For each vertex $v$, $\mathcal{N}(v)$ indicate the {\em neighborhood} of $v$, i.e. the set of nodes that are connected by an edge to $v$. The number of neighbors is the {\em degree} of $v$, denoted with $deg(v)$.

% A {\em vertex cover} of \( G \) is a subset \( C \subseteq V \) such that for every edge \( (u, v) \in E \), at least one of \( u \) or \( v \) belongs to \( C \). The {\em minimum weight vertex cover} problem seeks to find a vertex cover \( C \) that minimizes the total weight of the vertices in \( C \):  
% %
% %This problem can be formulated as the following optimization problem:  
% %
% \begin{mini}|s|
% {}{\sum_{v \in V} x_v \cdot w(v)}
% {}{}
% \addConstraint{x_u + x_v \geq 1 \hspace{1em} \forall (u, v) \in E}
% \addConstraint{x_v \in \{0, 1\} \hspace{1.5em} \forall v \in V}
% \label{opt:mainproblem}
% \end{mini}

% Here, \( x_v \in \{0, 1\} \) is a decision variable that indicates whether vertex \( v \) is included in the vertex cover (\( x_v = 1 \)) or not (\( x_v = 0 \)). The first constraint ensures that every edge \( (u, v) \in E \) is covered by at least one of its endpoints, and the objective function minimizes the total weight of the vertices included in the cover.  

\begin{figure*}
    \centering
    \includegraphics[width=0.85\textwidth]{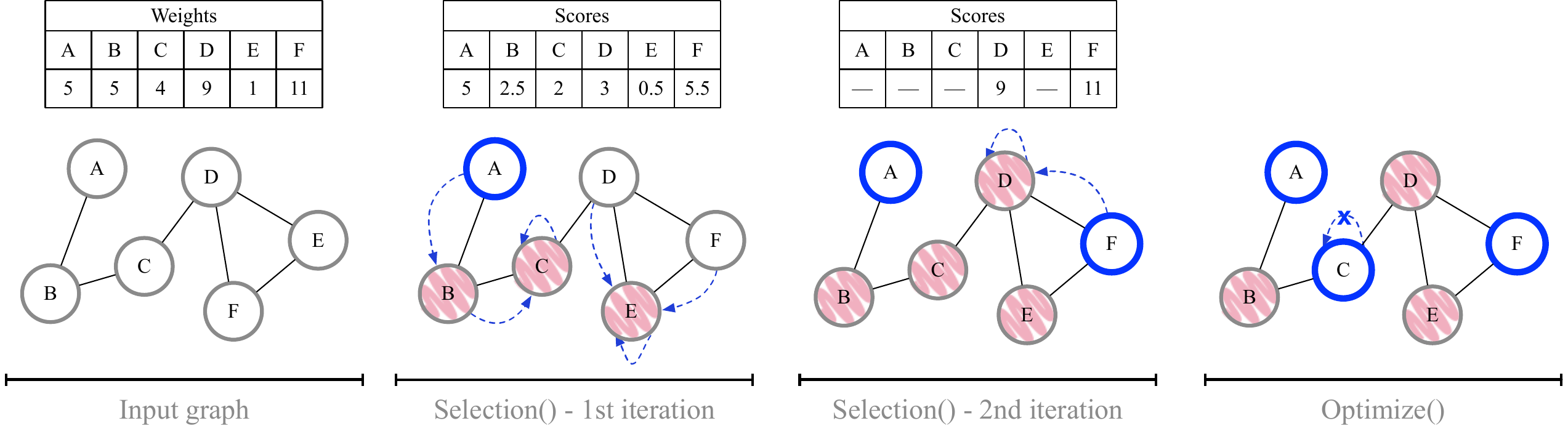}
    \caption{An example of \our in action. Colored vertex are in the cover set, while bold ones are inactive and not in the cover set. Dashed arrows indicate the $include(\cdot)$  action. Dashed marked arrows indicate the removal of a node}
    \label{fig:example}
\end{figure*}

%\subsection{The decentralized heuristic}\label{sec:dwmvc}
In the following, we provide a detailed description of our proposed solution, termed the Decentralized Weighted Minimum Vertex Cover (\our) algorithm. We consider an undirected graph \( G = (V, E) \), where each vertex \( i \in V \) has an associated positive {\em weight} defined by the function \( w: V \to \mathbb{R}^+ \). For each vertex $v$, $\mathcal{N}(v)$ indicates the {\em neighborhood} of $v$, and $deg(v)$ is its {\em degree}. A {\em vertex cover} of \( G \) is a subset \( C \subseteq V \) such that for every edge \( (u, v) \in E \), at least one of \( u \) or \( v \) belongs to \( C \). The {\em weighted minimum vertex cover} problem seeks to find a vertex cover \( C \) that minimizes the total weight of the vertices in \( C \).

%Each node in the graph follows the same set of procedures until its local condition indicates convergence. Therefore, our discussion adopts the perspective of a generic node $n$ within the network.
%
\our operates based on local coordination among neighboring nodes, borrowing the core underlying concept from the vertex-centric paradigm \cite{TLAV_survey}. \inote{here we can briefly describe the vertex-centric basic model}
Leveraging both its own data and information from neighboring nodes, each node autonomously evaluates its role within the vertex cover, while also participating in the determination of the status of its neighbors. Therefore, in the following, our discussion adopts the perspective of a generic node $n$ within the network.

% Our solution is divided into two main phases. In the first phase, called \selection (Algorithm \ref{alg:selection}), nodes communicate and collaborate to determine whether they should be inside the cover set or not. The second phase, called \optimize (Algorithm \ref{alg:optimize}), allows them to optimise the final cover. Nodes understand whether they should drop the cover, since their role is redundant. Auxiliary functions are employed to execute necessary actions in both phases. The behaviour of these functions, being straightforward service operations, is explicitly described within the text.

A high-level overview of \our, illustrating how \selection and \optimize are combined, is presented in Algorithm \ref{alg:dwvc}. Essentially, each node performs \selection iterations until its neighbors and itself are not settled. A node is settled when it is either in the cover or all its neighbors are in the cover.
\begin{algorithm}[ht!] 
\small
\caption{\our} 
\label{alg:dwvc} 
\begin{algorithmic}[1]
\State Each generic node $n$:
\State

\While{ $\exists u \in \mathcal{N}(n) \cup \{ n \} , \lnot u.isSettled$}
    \State \selection
\EndWhile
%\Until{$n$ is $settled$ and all $\mathcal{N}(n)$ is settled}
\State \optimize

\end{algorithmic}
\end{algorithm}
The primary algorithmic steps of the \selection are delineated in Algorithm~\ref{alg:selection}. We assume each node is characterized by a unique ID, where IDs are positive integers. Each node $n$ also has a local flag $n.\text{\it inCover}$ that highlights whether it should consider itself part of the cover. This flag is initially set at $False$. Moreover, to derive the behavior of nodes, we leverage a scoring function whose rationale is partially inspired by the scoring function defined in~\cite{CAI201964}. Specifically, a node $n$ that is outside the cover can compute its score relying on its $gain$ value, defined as: $gain(n) = |\{u \in \mathcal{N}(n)~|~u.inCover = False\}|$.
%
%\begin{equation}
%gain(n) = |\{u \in \mathcal{N}(n)~|~u.inCover = False\}|
%\end{equation}
%
Essentially, the $gain$ of $n$ measures how many other uncovered nodes would be covered if $n$ would enter the cover set. 
Note that, initially, the $gain$ corresponds to the degree of a node. However, the $gain$ value alone is not enough to evaluate the suitability of $n$  to be included in the cover, since weights are not considered. To this end, the final $score$ is computed in the following way:
\begin{dmath}
score(n) = \frac{w(n)}{gain(n)}
\end{dmath}
%
%It is easy to see that 
The $score$ function is minimised when a node has a lower weight and it covers a high number of other vertices. The $score$ of a node is then a robust local indicator for inclusion in the cover.
%good local indicator about the potential advantage to include the node in the cover. 
Specifically, The $score$ value is used by nodes during interactions to determine their respective fitness to enter the cover.

\begin{algorithm}[ht!]
\small
\caption{\selection } 
\label{alg:selection} 
\begin{algorithmic}[1]
\State Each generic node $n$:
\While{$\lnot n.inCover$ {\bf and} $\exists u \in \mathcal{N}(n), \lnot u.inCover$}
    \State $\mathcal{N^{\prime}} = \{m \in \mathcal{N}(n)~|~\lnot m.inCover\}$
    \State Get updated $score$ values from each $u \in \mathcal{N^{\prime}} $
    \State $\mathcal{N^{\prime\prime}} =  \mathcal{N^{\prime}} \cup \{ n \} $
    %\State Be $v^{\star} \in \mathcal{N^{\prime}}$ s. t. $ value(v^{\star})= \min\limits_{v \in \mathcal{N^{\prime}}} value(v)$
    \State $v^{\star} = \argmin\limits_{v \in \mathcal{N^{\prime\prime}}} score(v)$
    \State $include(v^{\star})$
\EndWhile
\State $settled(n.ID)$

\end{algorithmic}
\end{algorithm}

Given these preliminaries, \selection proceeds in the following way. A node $n$ starts executing a cycle, and it will continue to execute the same cycle until either it is in the cover set ($n.inCover = True$) or all its neighbors are in the cover (line 1 of Algorithm~\ref{alg:selection}). 

%\section{Algorithm}\label{sec:algorithm}

Within the cycle, $n$ considers the subset $N^{\prime}$ of neighbors that are outside the cover. It asks them for updated values of their $score$ (lines 2--3). In doing this, we assume it also sends them its own $score$. Node $n$ selects the vertex $v^\star$ that has the lowest $score$ among itself and the nodes in $N^{\prime}$ (lines 4-5). Then, $n$ communicates to $v^\star$ that should consider itself part of the cover, calling the function $include(\cdot)$ (line 6). The calling of the function involves the fact that, upon receiving this message, $v^\star$ sets its flag $v^\star.inCover$ to $True$ and send a message to its neighborhood announcing that it is now part of the cover set. If  $v^\star$ is $n$ itself, it simply sets its flag to $True$ and send the related communication to its neighbors. We assume that, in case the lowest score is shared among two or more nodes, the node with the highest ID is unequivocally selected. 
Disseminating information about a node being part of the cover is crucial, as it enables its neighbors to make informed decisions during this phase.

After ending the cycle, $n$ communicates to its neighbors that it has ended the first phase by sending a message to $\mathcal{N}(n)$ calling the function $settled$ with its own $n.ID$ as a parameter (line 8). 

\begin{algorithm}[ht!] 
\small
\caption{\optimize} 
\label{alg:optimize} 
\begin{algorithmic}[1]

\State Each generic node $n$:
\State $S = \{u \in \mathcal{N}(n)~|~u.inCover = True \} $
\If{$n.inCover$ {\bf and} $|S| = deg(n)$}
    \State $R = \emptyset$
    \ForAll{$m \in S$}
        \State $a = communicateDrop(n.ID)$
        \State $R = R \cup \{a\}$
    \EndFor
    %\If{$a \geq 0$ {\bf and} $a < n.ID$}
    \If{$\forall i \in R, i = -1$ {\bf or} $ \exists i \in R$ s.t. $i > n.ID$}
        \State $n.inCover = False$
    \Else{}
        \State $revokeDrop(n.ID)$
    \EndIf
\EndIf

\end{algorithmic}
\end{algorithm}

A node $n$ will not start executing the steps delineated in \optimize until itself has not ended the first phase and it has not received a communication from all of its neighbors that they also have ended the steps of \selection. This fact allows $n$ to consider the situation of its neighborhood as stable.

\begin{table*}[]
\footnotesize
\caption{Results with synthetic sparse graphs. Each value is averaged over all the experiments with the indicated topology. \timemetric is in milliseconds.}
\label{tab:synth_sparse}
\resizebox{\textwidth}{!}
{
\begin{tabular}{llrrrrrrrrrrrrrr}
\toprule
&
\multirow{2}{*}{topology} & 
\multicolumn{3}{c}{\ad} &
\multicolumn{9}{c}{$|V|$} &
\multicolumn{2}{c}{weight distribution} \\
\cmidrule(lr){3-5}
\cmidrule(lr){6-14}
\cmidrule(lr){15-16}

& & 5 & 10 & 15 & 64 & 128 & 256 & 512 & 1024 & 2048 & 4096 & 8192 & 16384 & uniform & power \\
\midrule
\multirow{3}{*}{\apw} 
& \nws & 0.63 & 0.73 & 0.79 & 0.71 & 0.71 & 0.72 & 0.72 & 0.72 & 0.72 & 0.72 & 0.72 & 0.72 & 0.73 & 0.70 \\
& \randg & 0.54 & 0.67 & 0.74 & 0.67 & 0.65 & 0.65 & 0.65 & 0.65 & 0.65 & 0.65 & 0.65 & 0.65 & 0.66 & 0.64 \\
& \barabasi & 0.43 & 0.57 & 0.63 & 0.58 & 0.54 & 0.55 & 0.54 & 0.54 & 0.54 & 0.53 & 0.54 & 0.54 & 0.55 & 0.54 \\
\midrule
\multirow{3}{*}{rounds} 
& \nws & 3.23 & 3.78 & 3.91 & 3.18 & 3.37 & 3.55 & 3.68 & 3.67 & 3.75 & 3.78 & 3.87 & 3.92 & 3.61 & 3.67 \\
& \randg & 3.14 & 3.63 & 3.81 & 3.13 & 3.18 & 3.35 & 3.47 & 3.58 & 3.72 & 3.70 & 3.75 & 3.88 & 3.49 & 3.57 \\
& \barabasi & 2.92 & 3.17 & 3.42 & 2.85 & 2.90 & 3.03 & 3.12 & 3.08 & 3.28 & 3.30 & 3.42 & 3.57 & 3.12 & 3.23 \\
\midrule
\multirow{3}{*}{\mpn}
& \nws & 9.87 & 18.06 & 25.79 & 16.98 & 17.62 & 17.80 & 17.99 & 17.99 & 18.08 & 18.13 & 18.24 & 18.32 & 17.81 & 18.00 \\
& \randg & 8.93 & 16.98 & 24.78 & 15.88 & 16.22 & 16.43 & 16.84 & 17.18 & 17.41 & 17.24 & 17.37 & 17.52 & 16.81 & 16.99 \\
& \barabasi & 7.28 & 13.25 & 18.95 & 12.74 & 12.59 & 12.97 & 12.95 & 12.83 & 13.35 & 13.41 & 13.63 & 13.97 & 12.98 & 13.33 \\
\midrule
\multirow{3}{*}{\timemetric}
& \nws & 20.46 & 35.32 & 51.96 & 0.53 & 1.04 & 2.08 & 4.48 & 9.25 & 18.97 & 37.62 & 75.92 & 173.33 & 36.21 & 35.62 \\
& \randg & 19.93 & 33.97 & 51.57 & 0.50 & 0.95 & 1.93 & 4.16 & 8.85 & 18.32 & 37.71 & 72.75 & 171.22 & 34.97 & 35.33 \\
& \barabasi & 16.22 & 24.76 & 34.66 & 0.38 & 0.74 & 1.47 & 3.06 & 6.14 & 12.68 & 26.44 & 54.28 & 121.73 & 25.13 & 25.30 \\
\bottomrule
\end{tabular}
}
\end{table*}

The \optimize (Algorithm~\ref{alg:optimize}) is used to reduce the number of nodes that are part of the cover, and hence to reduce the weight of the final solution, while maintaining the correctness of the solution. 
Specifically, it is executed only if a node $n$ is in the cover. If this is the case, node $n$ checks whether its entire neighborhood is also included in the cover (lines 1--2 of Algorithm~\ref{alg:optimize}). In this case, $n$ contemplates the possibility of withdrawing itself from the cover, since its presence is redundant. However, prior to finalizing this decision, $n$ initiates communication with its neighborhood, informing of its decision using function $communicateDrop$ and appending its ID to the message (lines 4--7). As a reaction, upon receiving the message, neighbors who intend to {\em remain} in the cover respond with a value of $-1$. Conversely, if a neighbor $m$ is contemplating the same action as $n$ (i.e., withdrawing the cover), a conflict may arise. To mitigate such conflicts, we establish a protocol wherein the node with the lower ID between the two opts to exit the cover. As node $m$ receives the ID of $n$ within the message, it can make an informed decision. It responds with its own ID, enabling $n$ to execute the appropriate action. In case all the conditions are met, $n$ eventually leaves the cover (lines 8--10). If the neighbors of $n$ do not receive further communications, they correctly assume that $n$ has exit the cover. In case $n$ decides to remain in the cover set, it sends a message using the $revokeDrop$ function with its own ID, announcing that its previous intention to leave the cover is forfeited. 
The final vertex cover is made up of all the nodes that have their $inCover$ flag set to $True$.

An example of \our in action is depicted in Figure \ref{fig:example}. 
During the first iteration of \selection,
these include messages are sent accordion to the score of the nodes: $\{A \rightarrow B, B \rightarrow C, C \rightarrow D, D \rightarrow E, E \rightarrow E, F \rightarrow E\}$.
At the end of the first iteration of the \selection, the cover is composed of $\{B, C, E\}$, a collection of nodes with low weight and well connected. During the second iteration, only $\{D, F\}$ are active, and $D$ joins the cover set due to its lower weight with respect to $F$. During the \optimize, $C$ realizes that all its neighbors are in the cover set, and then it removes itself from the cover set.

% \section{Mathematical Description as a Discrete-Time Evolutionary Process}\label{sec:math}

% \begin{dmath}
% p_n(t+1) = p_n(t) + \\ (1 - p_n(t)) \left( 1 - \frac{value_t(n)}{\sum\limits_{u \in \mathcal{N}(n)} (1 - p_u(t)) \cdot value_t(u)} \right) - p_n(t) \prod\limits_{u \in \mathcal{N}(n)} p_u(t)
% \end{dmath}

% \begin{equation}
% p_n(0) =  1 - \frac{value_0(n)}{\sum\limits_{u \in \mathcal{N}(n)} value_0(u)} 
% \end{equation}

% \begin{equation}
% value_0(n) = \frac{weight(n)}{deg(n)}
% \end{equation}

\section{Experimental Evaluation}
\label{sec:exp}

%This evaluation aims to assess \our performance in three criteria thoughtfully. 
The evaluation of \our was designed to seek an answer to the following questions:

\begin{itemize}
    \item \textit{How does \our behave in terms of convergence, and how many messages are exchanged by nodes? Do topologies have an impact?} This is discussed in Sections \ref{sec:convergence} and \ref{sec:messages}.

    \item \textit{Is the quality of the coverage provided by \our in line with centralized state-of-the-art approaches?} The quality of the solution is analyzed in Section \ref{sec:quality}.

    \item 
    Finally, our approach is also relatively fast when executed sequentially during the simulation. \textit{How does the execution time of \our adapts to different graphs?}  
    Even if this analysis doesn't adhere directly to the distributed nature of the approach, we believe it could be interesting for a potential multi-core implementation.
    We discuss this evaluation in Section \ref{sec:time}.

\end{itemize}

\subsubsection{Graph Datasets}
%We used a mix of synthetic and real graphs to conduct our evaluation. \\
%Synthetic graphs have been proven helpful in understanding the trends of the approach in a controlled setting, while real graphs provide a good understanding of actual on-field performances.\\

%\noindent
%\textbf{Synthetic Sparse Graphs.} 
We built a testbed of 164 undirected, node-weighted graphs with three topologies: \randg, \barabasi, and \nws, respectively created with the Erdős-Rényi, Barabási–Albert, and Newman–Watts–Strogatz strategies.
Random graphs are characterized by having roughly the same number of edges for each node (degree). In \barabasi graphs, also known as preferential attachment graphs, the degree distribution follows a power law, resulting in nodes with unusually high degrees compared to other nodes. Meanwhile, \nws graphs are characterized by a high clustering coefficient and short path lengths.
The order (number of nodes) of the graphs varied from $2^6$ to $2^{14}$, and their target average degree varies in the set $[5, 10, 15]$. 
The target average degree (\ad) was enforced by setting the probabilities for edge generation for each topology accordingly. 
%
%This way, the actual average degrees were very close to the target ones, with only negligible differences. 
%Hence, with the term average degree, and symbol \ad, we refer to the target average degree for generated graphs, and the real degree for the other graphs.
%
%Additionally, when the generated graph consisted of isolated components, we added an edge between these components to enforce a fully connected graph (nodes were chosen randomly between components). This adjustment was primarily applied to random graphs, and only a negligible number of edges were added.
%
Finally, we used two strategies to assign weights. In the first strategy, the weights of the nodes were drawn uniformly at random from the range $[20, 100]$, the same strategy employed in the work of Cai et al. \cite{CAI201964}, and Li et al. \cite{li2016efficient}.
For the second strategy, we use a power law distribution on the same interval $[20, 100]$ with the exponent fixed at $0.5$.
All the graphs were generated using the NetworkX library\footnote{\url{https://networkx.org/}}, then exported in a format suitable for \fastvc.
%\inote{add the other competitors in case}

%\noindent
%\textbf{Synthetic Dense Graphs.} 
% Similarly to the sparse graphs, we generated 32 denser graphs. We kept the number of nodes fixed at $2^{14}$, with the \ad as a fraction of the number of nodes in the set  $[0.02, 0.04, 0.06, 0.08, 0.10, 0.12, 0.14, 0.16, 0.18, 0.20]$. Topologies and weight distribution varied in the same way as the sparse graphs.
%
%To study dense graphs, we used the BHOSLIB\footnote{\url{https://networkrepository.com/bhoslib.php}} dataset, a popular benchmark set for testing the minimum vertex cover, widely used by the community (such as in \cite{wang2021fast} and \cite{cai2013numvc}). The dataset contains 35 instances of small but dense graphs (from 450 to 1534 nodes, from 185 to 684 average degrees) plus one very dense (\ad$=3712$) and relatively larger graph (frb100-40) of 4K nodes and around 7.5M edges.\\

%\noindent
%\textbf{Real Graphs.}
% \begin{figure}
%     \centering
%     \includegraphics[width=0.7\linewidth]{synth_rounds_order.pdf}
%     \caption{Convergence of \our when increasing the graph nodes and for different topologies}
%     \label{fig:synth_rounds_order}
% \end{figure}
%
%
The real graphs datasets consist of \numrealgraphs real graphs publicly available  \cite{nr} (see list in Table~\ref{tab:real}) . 
%The list of real graphs can be found in Table \ref{tab:real}.
%
The mix of chosen graphs is very heterogeneous in many aspects, ranging from some hundreds of nodes and edges to millions of nodes and edges. 
% Most notably, there are social network graphs (such as soc-youtube-snap, soc-FourSquare, socfb-Stanford3), web graphs (such as web-it-2004, web-uk-2005), interaction graphs (ia-reality, ia-fb-messages), biological graphs (bio-yeast, bio-diseasome), and others. The smallest graphs are composed of hundreds of nodes and edges, while the largest have millions of nodes and edges (for example, soc-livejournal has 4M nodes and 27M edges). The graph density is also very heterogeneous, with the average degree varying from a minimum of $2.26$ to a maximum of $181.19$.
In addition, the graph density varies from a minimum of $2.26$ to a maximum of $181.19$.
Since the original graphs are unweighted, we assigned node weights using the same uniform strategy we applied to the synthetic graphs.

%\noindent
\subsubsection{Simulation environment}
\our has been implemented as a Python script. All our evaluations are simulated, and each experimental value averages 10 independent runs.
All the simulations were executed on a single
workstation, an 8-core Intel i9-9900K @ 3.60GHz equipped with 64GB of RAM.

\subsection{Convergence}
\label{sec:convergence}

%\begin{table*}
%\centering
%\caption{Results with synthetic dense graphs from the BHOSLIB collection.}
%\label{tab:synth_dense}

%\resizebox{0.99\textwidth}{!}
%{
%\input{bhoslib_table.tex}
%}

%\end{table*}

%The purpose of this experiment is to 
We analyze the convergence and the network performance of \our in terms of the number of messages.
The convergence is measured in the number of iterations needed by the \selection plus one round of \optimize.
%

%%%%%%%%%%%%%%%%%%%%%%%%%%%
% ---------- CONVERGENCE
%%%%%%%%%%%%%%%%%%%%%%%%%%%%

With synthetic sparse graphs, the convergence is very similar across all experimental setups, ranging from a minimum of 2 to a maximum of 4 rounds. 
As we can observe in Table \ref{tab:synth_sparse}, %and Figure \ref{fig:synth_rounds_order}, 
the convergence speed increases slightly with the \ad and the number of nodes in the graphs. This trend is consistent for all the topologies. 
By comparing the topologies, it is evident that the \barabasi converges significantly faster at each order.
%
% In synthetic dense graphs (Table \ref{tab:synth_dense}), convergence takes more rounds than in the sparse graphs. Compared to the sparse graphs, the dense graphs have fewer nodes but a much higher \ad. This suggests that the density of the networks plays a significant role in the convergence speed. 
%
% However, the trend of the convergence speed grows very slightly with the \ad. For example, in the graph \textit{frb-30} (\ad$=369$), the convergence speed is around 6 rounds, while in \textit{frb-59}  (\ad$=1368$), an almost complete graph with 1540 nodes and \ad$\thicksim 1360$, the convergence speed is around 7 rounds. 
%
The results for the real graph are presented in Table \ref{tab:real}. 
The convergence is fast in every graph, with a max of 5 and a minimum of 3 rounds. As in the synthetic graphs, the sparsest graph tends to have a faster convergence, regardless of the number of nodes or edges.

\begin{figure*}
     \centering
     \begin{subfigure}[b]{0.48\textwidth}
         \centering
         \includegraphics[width=0.75\textwidth]{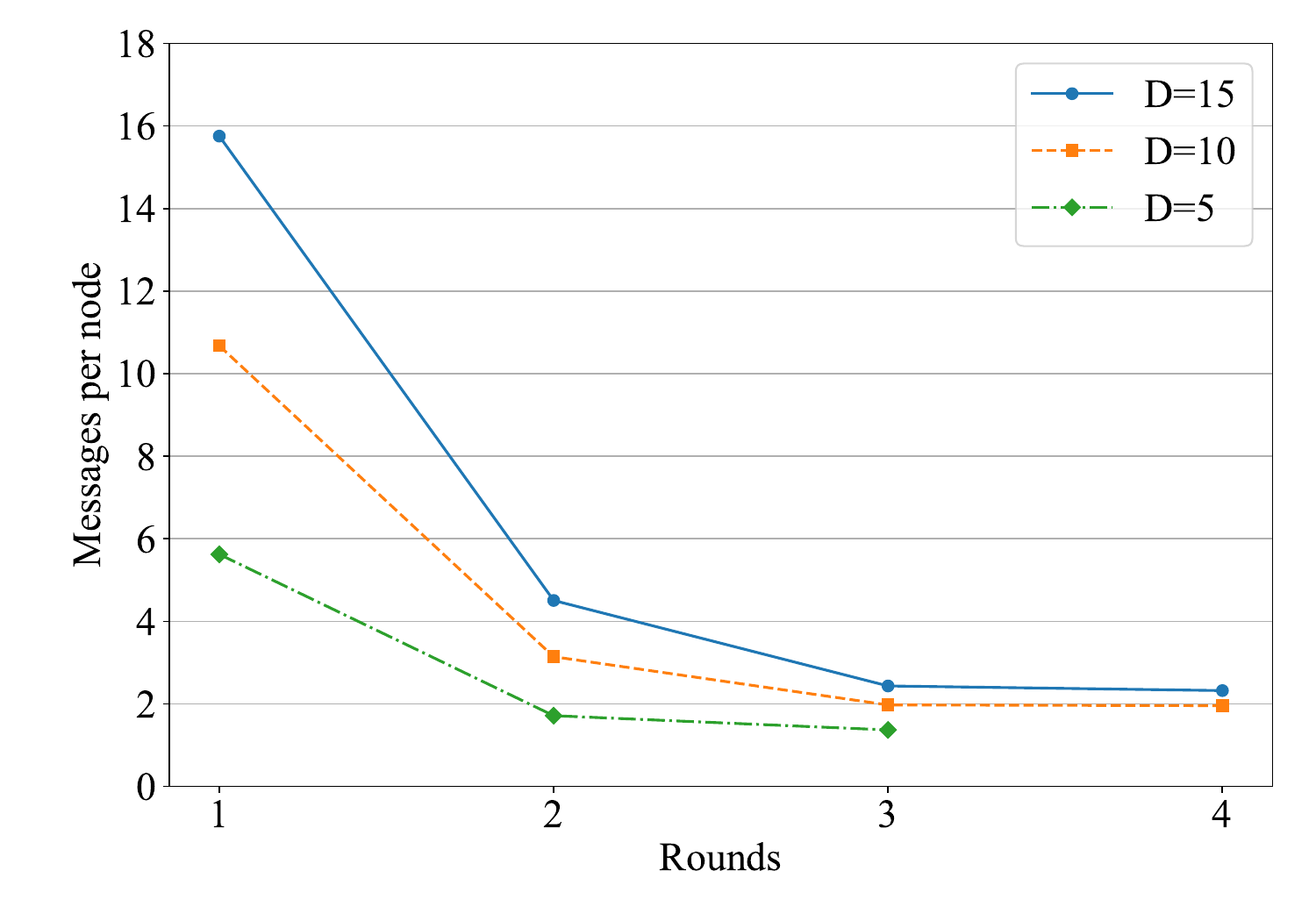}
         \caption{Varying \ad, \randg topology, $2^{14}$ nodes, uniform weights}
         \label{fig:synth_messages_degree}
     \end{subfigure}
     \hfill
     \begin{subfigure}[b]{0.48\textwidth}
         \centering
         \includegraphics[width=0.75\textwidth]{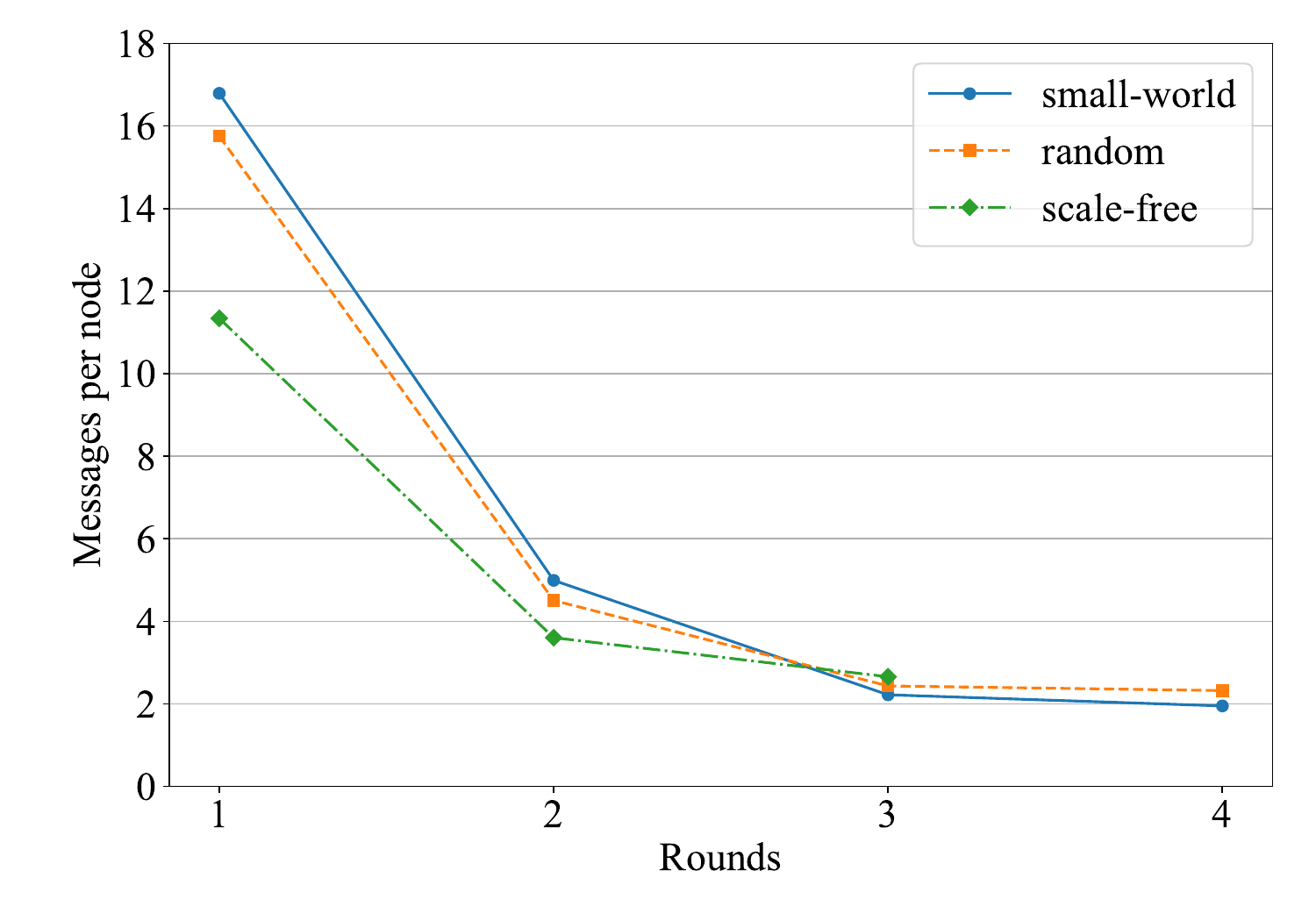}
         \caption{Varying topologies, $2^{14}$ nodes, uniform weights, \ad = 15}
         \label{fig:synth_messages_topologies}
     \end{subfigure}
        \caption{Messages per node}
        \label{fig:synth_messages}
\end{figure*}

%%%%%%%%%%%%%%%%%%%%%%%%%%%
% ---------- MESSAGES
%%%%%%%%%%%%%%%%%%%%%%%%%%%

\subsection{Messages}
\label{sec:messages}

% The number of messages is counted considering both the messages exchanged in the iterations of \selection and the execution of \optimize. 
We consider the number of messages per node (\mpn) by dividing the total messages by the number of nodes on the graph at the end of the simulation.
From the analysis of the synthetic sparse graphs, we can observe that, in general, \mpn drops with iterations.
This is due to the incremental nature of the approach, which inserts several nodes in the cover set during each iteration. Nodes already in the cover set are inactive during the \selection (see Algorithm \ref{alg:selection}), and a limited number of nodes is expected to be active in the \optimize phase (see Algorithm \ref{alg:optimize}).
Specifically, as seen in Figure \ref{fig:synth_messages}, the dropping trend is similar for different \ad (see Figure \ref{fig:synth_messages_degree}) and different topologies (see Figure \ref{fig:synth_messages_topologies}).
However, there is a difference in the number of messages per node (see Table \ref{tab:synth_sparse}): sparser networks seem to require fewer messages, and among topologies, the \barabasi is the one that requires fewer messages; the other two topologies are similar.
The same trend can be observed both in synthetic dense and real graphs.
The sparser the graph, the fewer messages per node. This confirms the results obtained with the sparse synthetic data and also depends on the fact that sparse graphs are generally observed to converge faster and, therefore, send fewer messages in total.
A notable exception is the real graph \textit{socfb-Stanford3} that, while its convergence is not the longest, is very dense and has the highest number of messages per node.

\subsection{Quality}
\label{sec:quality}

%%%%%%%%%%%%%%%%%%%%%%%%%%%
% ---------- APW
%%%%%%%%%%%%%%%%%%%%%%%%%%%

%The objective of the \problem is to find the set of nodes that covers all the edges and minimize the sum of their weight. To this end, 

To evaluate and compare the quality of a cover solution, we use the Average Percentage Weight (\apw), defined as the sum of the weights of the nodes in the cover set divided by the sum of the weights of all nodes in the graph, averaged for all the experiment runs.
Specifically, for a given graph $G=(V,E)$, we compute the following:
\begin{equation}
    APW = \frac{1}{r}\sum{\frac{\sum_{u \in C}w(u)}{\sum_{u \in V}w(u)}}
\end{equation}

where $C$ is the vertex cover and $r$ is the number of independent runs in the experiment (10, if not specified differently). 
When compared, lower values of \apw correspond to better solutions.

As a reference for the \apw, we compare \ours with \fastvc \cite{CAI201964} from Cai et al., a centralized state-of-the-art approach to solve the minimum vertex cover. \fastvc follows an incremental approach by finding an initial solution and then refines it iteratively until the cutoff time is reached.
%
%In typical usage, the cutoff is set to a reasonably high value to ensure that a good refinement is found.
%
We use \fastvc with a cutoff of \cutoff seconds, 
%the same as the original paper, 
to compute a good, albeit approximate, reference value.
%
%Similarly, we also tested with DynWVC \cite{cai2018improving}, still from Cai et al. The results were very similar to \fastvc, so we omitted them from the presentation in this paper.

%
%Conversely, we use a simple greedy heuristic as a baseline to provide an upper bound. The heuristics add one node at a time to the cover set by iterating over the following until all the edges are covered: selects one edge at random, adds the endpoint with the lowest weight to the cover set, and removes all the edges connected with the added node

%This experiment aims to evaluate the scalability in terms of \apw of \our with a set of generated sparse graphs. 
%

%\begin{figure*}
%     \centering
%     \begin{subfigure}[b]{0.3\textwidth}
%         \centering
%         \includegraphics[width=\textwidth]{synth_apw_size_d5.pdf}
%         \caption{\ad = 5}
%         \label{fig:y equals x}
%     \end{subfigure}
%     \hfill
%     \begin{subfigure}[b]{0.3\textwidth}
%         \centering
%         \includegraphics[width=\textwidth]{synth_apw_size_d10.pdf}
%         \caption{\ad = 10}
%         \label{fig:three sin x}
%     \end{subfigure}
%     \hfill
%     \begin{subfigure}[b]{0.3\textwidth}
%         \centering
%         \includegraphics[width=\textwidth]{synth_apw_size_d15.pdf}
%         \caption{\ad = 15}
%         \label{fig:five over x}
%     \end{subfigure}
%        \caption{\apw as a function of the graph order on synthetic sparse graphs, for different \ad and by varying the graph order. The solid line is \our, and the dotted line is \dyn. Lower is better.}
%        \label{fig:synth_apw_size}
%\end{figure*}

An overview of the \apw for the synthetic sparse graph can be found in Table \ref{tab:synth_sparse}.
%
%Also, in Figure \ref{fig:synth_apw_size} we observe the \apw for the three topologies, with different \ad and by varying the order of the graphs. 
%
The first observation we can make is that the \apw for both approaches is independent of the graph's order, when the density is fixed. 
%
%%% TO BE ADDED AFTER REVIEW
\inote{To be added after review. Interestingly, this also applies to the unweighted minimum vertex cover using different approaches \cite{sac25}.}
The \apw appears to be directly correlated with the \ad; that is, the higher the \ad, the higher the \apw. This is likely because, for the same graph order, a node can more easily cover 'lighter' edges
There is a noticeable distinction between the topologies in the \apw trend. The \nws topology appears to be the simplest to solve, followed by the \randg topology, and finally, the \barabasi topology.
%
% In general, as we can observe in Figure \ref{fig:synth_apw_size}, \fastvc outperforms \our for synthetics sparse graphs.
%
Finally, different weight distributions do not have an impact.

Regarding real graphs, the \apw of \our is practically on par with \fastvc, and for 5 of the largest graphs, such as soc-livejournal, it is even better. 
In detail, out of the \numrealgraphs graphs tested, \fastvc achieves results better than \our by more than 1\% in 10 graphs, and on average, the results found by \fastvc are only 0.8\% better.
%
% Similar observations can be made for dense synthetic graphs. For those, the difference with \fastvc is only 0.3\% on average, and \our is better for 6 graphs out of the 36 graphs, such as the \textit{frb59-26-4} and \textit{frb45-21-5} graphs.
%
\begin{table*}
\tiny
\centering
\caption{Results on real graphs, ordered by graph nodes. $|C|$ is the size of the cover set.}
\label{tab:real}
\resizebox{\textwidth}{!}
{
\begin{tabular}{lrrrrrrrrrr}
\toprule
\multirow{2}{*}{Graph} &
\multirow{2}{*}{$|V|$} &
\multirow{2}{*}{$|E|$} &
\multirow{2}{*}{\ad} &
\multicolumn{5}{c}{\our} &
\multicolumn{2}{c}{\fastvc} \\
\cmidrule(lr){5-9}
\cmidrule(lr){10-11}

&  &  & & \apw & $|C|$ & \mpn & rounds & \timemetric & \apw & \apw diff \\
\midrule
bio-diseasome & 516 & 1188 & 4.60 & 0.50 & 288.30 & 7.15 & 3.00 & 0.00 & 0.50 & -0.28\% \\
soc-wiki-Vote & 889 & 2914 & 6.56 & 0.44 & 424.50 & 7.92 & 3.00 & 0.00 & 0.43 & -2.67\% \\
ia-fb-messages & 1266 & 6451 & 10.19 & 0.45 & 600.20 & 10.41 & 3.00 & 0.01 & 0.43 & -2.7\% \\
%bio-yeast & 1458 & 1948 & 2.67 & 0.31 & 480.30 & 4.50 & 2.90 & 0.00 & 0.30 & -2.44\% \\
inf-power & 4941 & 6594 & 2.67 & 0.42 & 2331.70 & 5.10 & 3.00 & 0.02 & 0.41 & -2.83\% \\
ca-Erdos992 & 5095 & 7516 & 2.95 & 0.09 & 465.40 & 3.17 & 2.10 & 0.01 & 0.09 & -0.22\% \\
ia-reality & 6809 & 7680 & 2.26 & 0.01 & 81.00 & 2.29 & 2.00 & 0.01 & 0.01 & 0.0\% \\
%ca-HepPh & 11204 & 117619 & 21.00 & 0.55 & 6656.60 & 21.15 & 4.90 & 0.21 & 0.55 & -0.92\% \\
%web-indochina-04 & 11358 & 47606 & 8.38 & 0.60 & 7436.60 & 14.26 & 4.90 & 0.09 & 0.60 & -0.9\% \\
socfb-Stanford3 & 11586 & 568309 & 98.10 & 0.73 & 8690.30 & 78.50 & 4.00 & 1.64 & 0.72 & -1.32\% \\
%ca-AstroPh & 17903 & 196972 & 22.00 & 0.61 & 11632.50 & 23.12 & 4.80 & 0.28 & 0.60 & -0.71\% \\
ca-CondMath & 21363 & 91286 & 8.55 & 0.54 & 12693.70 & 11.76 & 4.00 & 0.15 & 0.54 & -0.8\% \\
tech-as-caida2007 & 26475 & 53381 & 4.03 & 0.13 & 3783.10 & 5.42 & 3.00 & 0.10 & 0.13 & -1.07\% \\
%soc-epinions & 26588 & 100120 & 7.53 & 0.35 & 10036.60 & 7.10 & 3.00 & 0.15 & 0.34 & -1.94\% \\
%ia-email-EU & 32430 & 54397 & 3.35 & 0.02 & 829.90 & 3.13 & 2.10 & 0.08 & 0.02 & -0.41\% \\
ia-enron-large & 33696 & 180811 & 10.73 & 0.35 & 12991.10 & 9.69 & 3.10 & 0.22 & 0.34 & -0.78\% \\
tech-p2p-gnutella & 62561 & 147878 & 4.73 & 0.25 & 15852.90 & 6.23 & 3.00 & 0.32 & 0.25 & -0.65\% \\
rec-amazon & 91813 & 125704 & 2.74 & 0.49 & 50354.80 & 5.58 & 3.10 & 0.50 & 0.47 & -2.91\% \\
web-uk-2005 & 129632 & 11744049 & 181.19 & 0.98 & 127774.00 & 283.94 & 7.00 & 80.55 & 0.98 & 0.03\% \\
%soc-douban & 154908 & 327162 & 4.22 & 0.06 & 8725.40 & 5.34 & 2.90 & 0.68 & 0.06 & -0.16\% \\
soc-gowalla & 196591 & 950327 & 9.67 & 0.41 & 86919.60 & 10.28 & 3.80 & 1.58 & 0.40 & -1.92\% \\
web-it-2004 & 509338 & 7178413 & 28.19 & 0.79 & 415054.90 & 47.28 & 6.90 & 38.08 & 0.79 & 0.16\% \\
%soc-flickr & 513969 & 3190452 & 12.41 & 0.28 & 157165.70 & 9.34 & 3.90 & 7.07 & 0.28 & -0.65\% \\
soc-FourSquare & 639014 & 3214986 & 10.06 & 0.14 & 91342.80 & 8.27 & 3.30 & 4.86 & 0.14 & -0.76\% \\
%soc-digg & 770799 & 5907132 & 15.33 & 0.13 & 106153.70 & 13.09 & 3.10 & 9.15 & 0.13 & -1.53\% \\
ca-hollywood-2009 & 1069126 & 56306653 & 105.33 & 0.77 & 865907.70 & 62.13 & 6.10 & 158.06 & 0.77 & 0.22\% \\
soc-youtube-snap & 1134890 & 2987624 & 5.27 & 0.23 & 285223.30 & 6.00 & 3.10 & 7.19 & 0.23 & 0.96\% \\
tech-as-skitter & 1694616 & 11094209 & 13.09 & 0.29 & 540528.30 & 19.96 & 4.10 & 19.48 & 0.30 & 2.26\% \\
soc-livejournal & 4033137 & 27933062 & 13.85 & 0.45 & 1920099.40 & 17.25 & 5.70 & 78.44 & 0.46 & 1.78\% \\
\bottomrule
\end{tabular}

}
\end{table*}

\subsection{Execution time}
\label{sec:time}

%\begin{figure}
%    \centering
%    \includegraphics[width=0.98\linewidth]{synth_dense_fit_exp.pdf}
%    \caption{Fitting of the running time with a quadratic polynomial}
%    \label{fig:fitting}
%\end{figure}

This section evaluated the running time (\timemetric) of \our when run as a sequential solver. 
%While this doesn't highlight any distributed feature, it can be useful to demonstrate the applicability of distributed approaches for sequential in-memory computation. In fact, using a distributed vision as a guiding principle can benefit the design of sequential or parallel in-memory computation programs by improving modularity, exploiting locality, and providing scalability and flexibility.
%
The \timemetric for \ours is taken by running the algorithm in-memory and using the python’s time package with the  \texttt{time.process_time()} method at the start and end of the simulation.
\timemetric is expressed in seconds unless it is specified otherwise.

With sparse synthetic graphs, see Table \ref{tab:synth_sparse}, time increases with higher order and average degree, while weight distribution has minimal impact. Vertex covers in scale-free networks are found significantly faster than small-world and random networks.
%
%Similar results can also be observed with dense synthetic graphs (Table \ref{tab:synth_dense}) and real graphs (Table \ref{tab:real}).
%
%In particular, for dense synthetic graphs, \timemetric fits a quadratic polynomial (see Figure \ref{fig:fitting}), with an R-squared test of $0.99$, which indicates a very good fit. The analysis of the polynomial indicates that there is a turning point at about \ad $= 380$ in which the trend increases, suggesting that beyond this point, the \timemetric grows more rapidly.
%
%
Finally, a direct comparison with \fastvc would be difficult due to its computation limited by the cutoff time, which we set at \cutoff seconds. However, we can indirectly observe that, for some graphs, \our obtains better cover sets in less than the cutoff case, as in the case of \textit{soc-youtube-snap}, \textit{tech-as-skitter}, and \textit{soc-livejournal} real graphs.
For example, in the case of the \textit{tech-as-skitter} real graph, \our finds a solution that is 2.26\% better than \fastvc in 19.48 seconds, compared to the \cutoff used by \fastvc.

% \subsection{Takeaways}

% Following the experimental evaluation, we can draw the following
% main points about \our:

% \begin{itemize}
% \item Its convergence speed is generally low for large graphs and depends on the graph's density. 
% We observed very fast convergence in sparse graphs, which increases with density but eventually plateaus, as the difference between dense and extremely dense graphs is minimal.

% \item The quality of the cover solution found by \our is generally very high, comparable to current centralized state-of-the-art approaches, and in some instances, especially in larger graphs, we even outperformed them.

% \item The topologies have a measurable impact on the performance of \our, particularly on convergence speed and execution time. Specifically, we observed better performance with the \barabasi topology compared to the \nws and \randg topologies.

% \item When used as an in-memory sequential solver, the execution time of \our seems dependent on the density of graphs. Notably, for some graph instances, \our finds better solutions faster than the centralized competitor.

% \end{itemize}

\section{Conclusion}
\label{sec:conclusion}

We presented a decentralized protocol for computing MWVC using only local communication and minimal coordination. The approach yields high-quality solutions across diverse topologies, with low overhead. Our results support its applicability in settings where scalability, resilience, and distributed control are essential. Future work will focus on asynchronous variants and integrating node churn.

% In this paper, we propose \our, a decentralized approach to solve the \problem problem. 
% %
% The algorithm relies on local node information, which makes it particularly suitable for application in distributed contexts, such as multi-agent and peer-to-peer networks, where entities can autonomously decide their inclusion in the vertex cover. 
% %
% Experimental results show fast convergence in many different graph topologies and settings, particularly in the \barabasi case, which makes it practical to use in multi-agent and computer networks.
% %
% Further, results highlight its competitive performance against a centralized solver, both in terms of solution quality and execution time when running as a sequential solver.
% %
% This last point opens the possibility of exploring a centralized implementation of \our for multi-core architecture as future work.

% speed, multi-core

%\section*{Acknowledgment}


\bibliographystyle{IEEEtran}
\bibliography{biblio}
%\thebibliography{biblio.bbl}

\end{document}